# Silicone incorporation into an esterquat based fabric softener in presence of guar polymers

Evdokia K. Oikonomou[1]\*, Camille Grandisson[1], Konstantin Golemanov[2], Ritu Ahuja[2] and Jean-François Berret[1]

[1]*Laboratoire Matière et Systèmes Complexes, UMR 7057 CNRS Université Denis Diderot Paris-VII, Bâtiment Condorcet, 10 rue Alice Domon et Léonie Duquet, 75205 Paris, France*
[2]*Solvay Research & Innovation Center Singapore, 1 Biopolis Drive, Amnios, Singapore 138622*

**Abstract**
Fabric softeners are widely used to make clothes soft and impart nice smell. Household conditioners are water-based formulations consisting of quaternary ammonium surfactants assembled in vesicles. On the other hand, formulations used for textiles finishing are typically silicone oil emulsions. Here, these technologies are combined in a novel industrial formulation which contains both surfactant vesicles and silicone oil. In particular, we insert an amino-modified silicone oil into a recently developed softener comprising vesicles of an esterquat and guar polymers. Two strategies based on the use of esterquat or guar gum as stabilizers for the silicone incorporation in the fabric softener are proposed. Dynamic light scattering (DLS), cryogenic transmission electron microscopy (cryo-TEM), optical and fluorescent microscopy are applied to study the stability and the insertion of the silicone oil into the softener. It is shown that silicone oil is stabilized by the surfactant and the guar gum in nano- or micro-metric droplets. The vesicles structure and properties are not affected by the silicone oil while the formulation's stability is preserved. These findings suggest that the proposed route for adding silicone oil into a topical conditioner is a promising route for multi-purpose fabric softener development.

# 1. Introduction

Fabric softeners, either household or those used for textiles finishing, are colloidal dispersions which impart softness to the fabric. Household fabric softeners are made up from double-chain cationic surfactants assembled into multi- or uni-lamellar vesicles. [1] The vesicles ensure formulation stability, acting also as vehicles for driving the surfactant bilayers and other water insoluble additives such as essential oils [2] or fragrance capsules [3] on the textiles. Recently, the vesicles were shown to adsorb onto the cellulose fibers [1,4,5,6,7] due to electrostatics, forming supported vesicular layers in water. [8] After drying, the vesicles are probably converted into supported bilayers on the fibers offering lubrication and softness while other mechanisms have been also proposed. [9,10]

There has been much progress on surfactants used in softeners formulations [11] while the replacement of dialkyldimethylammonium surfactants by esterquats is considered as a milestone. Esterquats are now dominant in the market thanks to their affordable cost and biodegrability since they bear hydrolysable ester groups. Nevertheless, nowadays further improvement of the fabric conditioners is ongoing due to the consumers' demands, market competition and sustainability concerns. In this framework, the palm oil origin of esterquats [12,13] is now under review due to the massive palm oil consumption which receives criticism. [14] Thus, apart from novel esterquats, [13] the incorporation of natural ingredients is under evaluation. [15,16] A sophisticated strategy was recently proposed by Solvay based on a significant reduction (> 50 wt. %) of the esterquats content in concentrated conditioners. Derived from the seeds of the plant cyamopsis tetragonalobus (guar gum), a cationic guar gum (C-Guar) known to impart softness on hair [17,18] and a hydroxypropyl-guar (HP-Guar) typically used as thickener, [19] were added in minute quantities to compensate the surfactant reduction. Indeed, improved softness and fragrance delivery were identified. [20,21] Although the polymers changed the microscopic structure of the formulation, it was shown that the bilayer characteristics, the vesicles properties and the rheological characteristics compared to the benchmark formulation were maintained. [4] In this work, we intent to further improve this vesicles/guar formulation by introducing an additional component, an amino-silicone oil in small quantities. Introducing an additive into industrial formulations is often delicate as the physicochemical properties should remain unmodified. However, the impact of additives such as water insoluble oils on the softener vesicles has been scarcely discussed in the literature. [2,16,22]

Silicone oils are typically utilized as finishing textile softeners, the most important textile chemical after-treatment. [23] They confer soft handle, smoothness, flexibility, pliability, antistatic properties, water repellency and sewability making the cloths attractive to the clients. Silicone combines hydrophobic properties with low glass transition temperature, elasticity, low surface tension, film forming ability and chain flexibility. The latter derives from the rotational freedom of the Si-O-Si linkages and the low interaction energies between the methyl groups, enabling the atoms to obtain the minimal energy structure at interfaces with lower surface pressure than the organic polymers. [23] Chemically modified silicones with groups such as aminoalkyls attached to the poly(dimethyl siloxane) backbone bring the so-called 'supersoft' handle. [23,24,25] These premium-grade softeners present enhanced adsorption and durability thanks to the cationic charges. [26] These silicones form a semi-interpenetrated network between the fibers based on electrostatic forces, crosslinking and H-bonds. [24,27] Silicone softeners are prepared in nano-, micro- or macro-emulsion based on the oil droplet size which is crucial for the performance. [28,29,30] Deeper penetration in the fiber, but lower bending ability and abrasion resistance is obtained by nano-emulsions. [28,31,32,33] To compensate these trends, mixtures of nano-, micro- and macro-emulsions can be used. [28] The choice of the surfactants used as emulsifiers for the droplets formation [34] is determinant for the droplet size, stability and performance. [28] Ionic, non-ionic surfactants or mixtures of them are generally used but non-ionic ones seem to be more efficient. [25,27,35,36] However, both environmental and performance issues lead to research for improved silicone softeners. For instance, hydrophilic-modified silicone polymers [37,38] can face up water repellency while silicone analogues of quats, the so-called silicone-quats have been developed. [23,28]

Here, we attempt to combine both esterquat and amino-modified silicone oil, in order to benefit of their different superior properties, in a novel formulation with low esterquat content. We thus incorporated low amounts of amino-modified silicone on the esterquat/guars containing softener described before. Knowing that the incorporation of additional ingredients into a formulation is often cumbersome, we avoided surfactant emulsifiers. Instead, we used either esterquat (bilayers/vesicles) or HP-Guar, already contained in the softener for stabilizing silicone. To the best of our knowledge this strategy has not been applied before for silicone conditioners. Double-chain surfactants such as phospholipids have been already reported as emulsifier agents for water-in-oil [39,40,41] or oil-in-water emulsions. [42,43] The phospholipids are considered to be embedded in the interfacial film [44] lowering the interfacial tension as shown also for Janus olive/silicone oil droplets. [43] Nonetheless, the interaction mechanisms of vesicles with the oil−water interfaces remain unclear. Recently, a mechanism of dynamic vesicles modification upon adsorption on oil droplets was proposed by Kataoka-Hamai et al. [45] Vesicles can be ruptured exposing its hydrophobic tails of lipids to the oil surface due to their high hydrophobic nature. The deformed vesicles therefore consist of an unbound bilayer and a bound monolayer. Depending on the interactions, sometimes they are deformed dynamically, encapsulating internal membrane structures, or desorbed from the interface. Under different conditions, liposomes were recently reported as stabilizers forming oil-in-water Pickering emulsion upon adsorption on the interface with oils such as sunflower [46] or in all aqueous emulsions. [47] As far as the HP-Guar is considered, polysaccharides were recently considered as stabilizers by forming a thick, gel-like steric barrier around the oil droplets, decreasing mobility and preventing them from coalescing [48] although they are generally used in presence of proteins. [49]

The aim of this article is to propose methodologies for the incorporation of silicone oil without any detrimental effects for the softener's properties. To this end, silicon oil is added in an esterquat/guars containing softener without using additional emulsifier surfactants. In particular, we explored the possibility of using the esterquat vesicles or HP-Guar of the softener as stabilizers. Both strategies resulted in good amino-silicone oil dispersion into the new formulation showing that this approach could be applied in other applications too. Physicochemical studies defined the impact of the silicone insertion on the vesicles properties and the softener characteristics. The developed formulations were shown to be stable upon time while the silicone droplets were characterized by dynamic light scattering (DLS), optical microscopy and cryo-TEM.

## 2. Experimental Section
### 2.1. Materials and sample preparation
**Materials.** Silicone oil, 2-3% aminopropylmethylsiloxane – dimethylsiloxane copolymer (80-120 cst, Mw 4500-6000) abbreviated as AMPS was purchased by Gelest. Its chemical structure is provided in Figure 1a. The polysaccharide polymers are a cationic guar gum (C-Guar, molecular weight $0.5 \times 10^6$ g mol$^{-1}$) and a hydroxypropyl guar gum (HP-Guar, molecular weight $2 \times 10^6$ g mol$^{-1}$), both synthesized by Solvay® (Figures 1b and 1c respectively). The second-order autocorrelation functions $g^{(2)}(t)$ of filtered C-Guar and HP-Guar dispersions at 0.02 wt. % are shown as a function of delay time in **Supporting Information, Figures S1a and S1b** respectively. Their hydrodynamic diameters (50 and 350 nm respectively) are larger than those expected from polymers of molecular weight $M_W$ = 0.5 and $2 \times 10^6$ g mol$^{-1}$ respectively, sug-gesting that the chains may be associated in water and form hydrocolloid particles. [50] Electrophoretic mobility measurements on dilute dispersions give zeta potential of +30 mV and 0 mV respectively. The esterquat surfactant ethanaminium, 2-hydroxyN,N-bis(2-hydroxyethyl)-N-methyl-esters with saturated and unsaturated C16-18 aliphatic chains, abbreviated TEQ in the following, was provided by Solvay®. The gel-to-fluid transition related to the long-range order of the surfactant molecules in the membrane is found at $T_M$ = 60 °C. The counterions associated to the quaternized amines are methyl sulfate anions. TEQ chemical structure is presented in Figure 1d along with its characterization by optical microscopy and cryo-TEM. Optical microscopy at 1 wt. % show well-contrasted objects, spherical (yellow arrows) or faceted (red arrows) identified as micro-sized vesicles. This latter phenomenon is related to mechanical stresses generated during preparation and the liquid-crystalline order of the alky chains within the layer, as shown previously by wide-angle x-ray scattering. [51] Cryo-TEM shows 100-300 nm unilamelar and multi-vesicular vesicles indicated by yellow and red arrows respectively. A more detailed characterization is presented in our previous work. [1] The fluorescent dye PKH67 was purchased by Aldrich. The water used for all formulations and dilutions was MilliQ water.

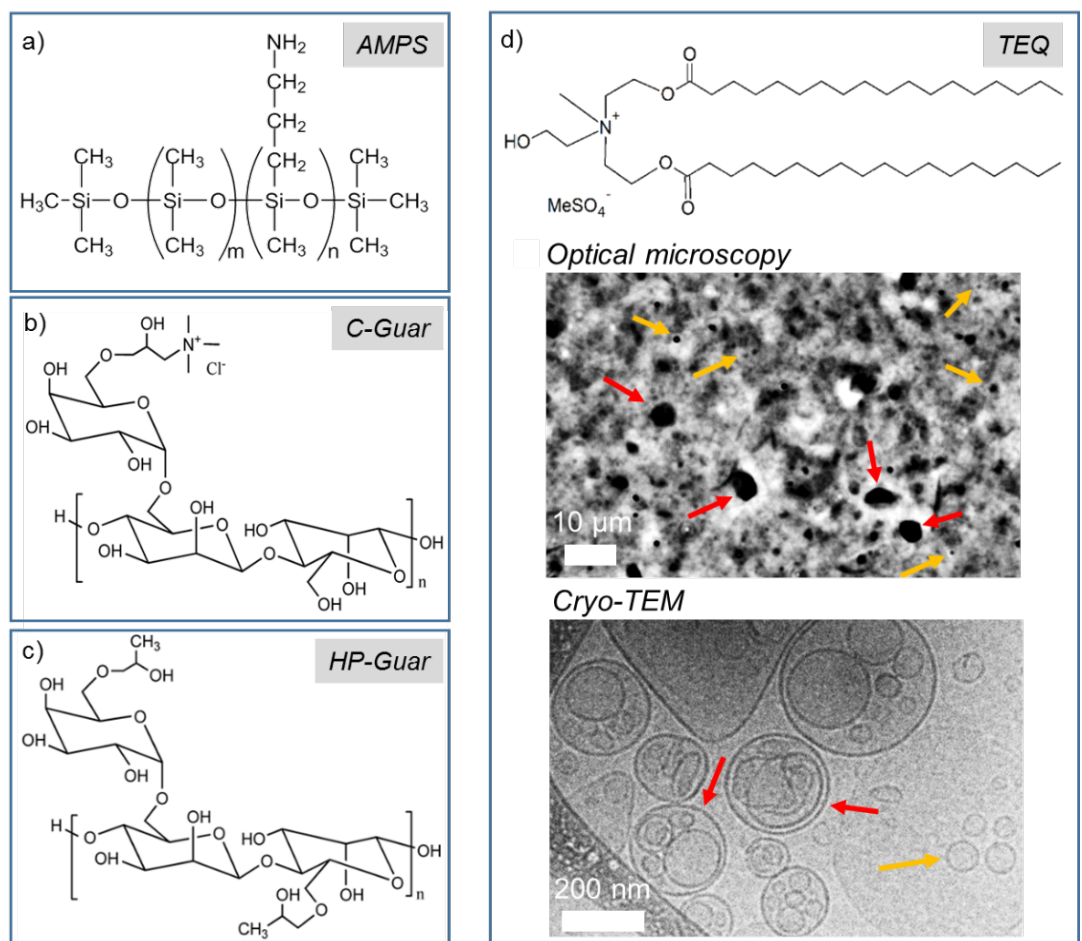

*Figure 1.* Chemical structures of (a) AMPS, (b) C-Guar, (c) HP-Guar and (d) TEQ. Optical microscopy and cryo-TEM images of 1 wt. % TEQ aqueous dispersion are also given in (d). The vesicles are indicated by red and yellow arrows.

**Sample preparation**

For the preparation of the new softener formulation containing amino-modified silicone oil (AMPS), esterquat TEQ and guar polymers at the following concentrations: TEQ 6 wt. %/AMPS 1 wt. %/ C-Guar 0.2 wt. % /HP-Guar 0.4 wt. %, two different strategies were followed. Prior to use, TEQ was first melted at 60 °C.

*Strategy I: TEQ used as oil dispersant*

Formulation A - TEQ emulsified AMPS abbreviated as *AMPS$_{TEQ}$*. 20 g AMPS aqueous dispersion was placed at an Ultra-Turrax T-18 (IKA) mixer. The speed was set at 25,000 rpm. After one minute, 2 g melted TEQ were added. Blending continued for 10 min. Final concentrations: AMPS 20 wt. %, TEQ 2 wt. %.

Formulation B - Softener 1 abbreviated as *SOFT 1*. C-Guar and HP-Guar were dispersed at 60°C water under stirring. Melted TEQ was then added dropwise. Finally, adequate quantity of AMPS$_{TEQ}$ was added at the formulation to obtain AMPS$_{TEQ}$ 1 wt. %. Final concentrations: TEQ 6 wt. %/AMPS$_{TEQ}$ 1 wt. %/ C-Guar 0.2 wt. % /HP-Guar 0.4 wt. %. The experimental process is schematically represented in Scheme 1a.

*Strategy II: HP-Guar used as oil dispersant*

Formulation C - Softener 2 abbreviated as *SOFT 2*. A dispersion of AMPS emulsified by HP-Guar abbreviated as *AMPS$_{HP-Guar}$* was fist prepared by mixing 1 wt. % AMPS with 0.4 wt. % HP-Guar for 10 min at an Ultraturax blender (speed 25,000 rpm). This dispersion was then heated at 60°C. C-Guar and melted TEQ were added at the formulation (final concentrations: 0.2 and 6 wt. % respectively). Final obtained concentrations: TEQ 6 wt. %/AMPS$_{HP-Guar}$ 1 wt. %/ C-Guar 0.2 wt. % /HP-Guar 0.4 wt. %. The experimental process is schematically represented in Scheme 1b.

An AMPS$_{HP-Guar}$ dispersion containing 1 wt. % AMPS and 0.8 wt. % AMPS was also prepared for comparison reasons but was not further used.

It has to be underlined that both procedures, using either TEQ surfactant or HP-Guar as amino-silicone dispersants result in the same softener formulation. SOFT 1 and SOFT 2 contain identical quantities of all components.

The new softeners were proven to present softening ability and stability as shown in the patent WO 2020/125949 A1.

*Reference formulations*

Formulation D: A formulation containing TEQ 6 wt. % + AMPS$_{TEQ}$ 1 wt. % was prepared for comparison reasons by adding AMPS$_{TEQ}$ in a 6 wt. % TEQ at 60°C.

Formulation E: A reference softener containing TEQ 6 wt. %/ C-Guar 0.2 wt. % /HP-Guar 0.4 wt. % was formulated as described elsewhere [4] for comparison reasons. The properties of this formulation are described in the related patent. [21]

The pH was adjusted at 4.5 for all softeners after the guars addition and at the final step. The concentrations mentioned in the manuscript for AMPS$_{TEQ}$ and AMPS$_{HP-Guar}$ refer to the AMPS concentration.

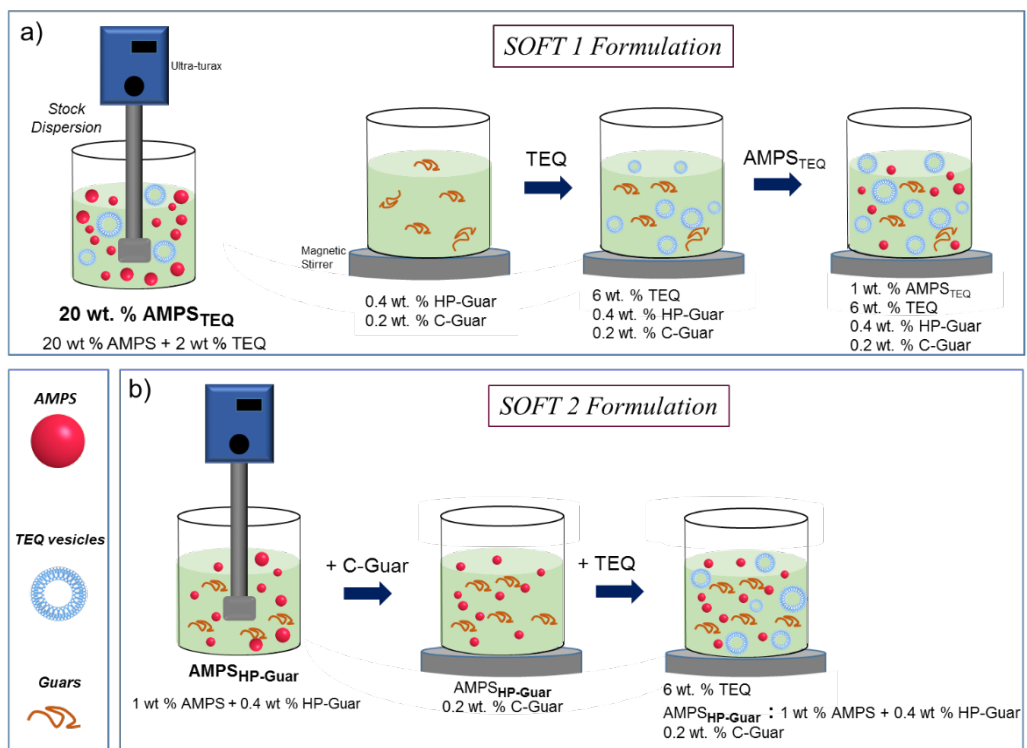

**Scheme 1.** *Schematic representation of the two different experimental processes followed for the formulation of (a) SOFT 1 and (b) SOFT2.*

## 2.2. Experimental techniques

**Light scattering.** The scattering intensity $I_S$ and hydrodynamic diameter $D_H$ were measured using the Zetasizer Nano ZS spectrometer (Malvern Instruments, Worcestershore, UK). A 4 mW He−Ne laser beam ($\lambda$ = 633 nm) is used to illuminate the sample dispersion, and the scattered intensity is collected at a scattering angle of 173°. The Rayleigh ratio $\mathcal{R}$ was derived from the intensity according to the relationship: $\mathcal{R} = (I_S - I_w)n_0^2 \mathcal{R}_T / I_T n_T^2$ where $I_w$ and $I_T$ are the water and toluene scattering intensities respectively, $n_0$ = 1.333 and $n_T$ = 1.497 the solution and toluene refractive indexes, and $\mathcal{R}_T$ the toluene Rayleigh ratio at $\lambda$ = 633 nm ($\mathcal{R}_T$ = $1.352 \times 10^{-5}\ cm^{-1}$). The second-order autocorrelation function $g^{(2)}(t)$ is analyzed using the cumulant and CONTIN algorithms to determine the average diffusion coefficient $D_C$ of the scatterers. The hydrodynamic diameter is then calculated according to the Stokes-Einstein relation, $D_H = k_B T / 3\pi \eta D_C$, where $k_B$ is the Boltzmann constant, $T$ the temperature and $\eta$ the solvent viscosity. Measurements were performed in triplicate at 25 °C after an equilibration time of 120 s.

**Zeta Potential**. Laser Doppler velocimetry (Zetasizer, Malvern Instruments, Worcestershore, UK) using the phase analysis light scattering mode and detection at an angle of 16° was performed to determine the electrophoretic mobility and zeta potential of the different dispersions studied. Measurements were performed in triplicate at 25 °C, after 120 s of thermal equilibration.

**Optical Microscopy**. Phase-contrast and bright field images were acquired on an IX73 inverted microscope (Olympus) equipped with 20×, 40×, and 60× objectives. Seven microliters of the dispersion were deposited on a glass plate and sealed into a Gene Frame (Abgene/ Advanced

Biotech) dual adhesive system. An Exi-Blue camera (QImaging) and Metaview software (Universal Imaging Inc.) were used as the acquisition system.

**Fluorescent Microscopy**. For fluorescent microscopy, 3 µl PKH67 were added at 97 µl AMPS$_{HP-Guar}$ 1 wt. % and were vortexed rapidly. PKH67 is used in cellular biology as a green fluorescent molecular linker developed for cell membrane labeling. It is characterized by an absorption maximum at 490 nm and an excitation maximum at 502 nm.

**Cryogenic transmission electron microscopy (Cryo-TEM)**. A few microliters of the samples were deposited on a lacey carbon coated 200 meshes (Ted Pella). The drop was blotted with a filter paper on a VitrobotTM (FEI) and the grid was quenched rapidly in liquid ethane, cooled with liquid nitrogen, to avoid the crystallization of the aqueous phase. The membrane was finally transferred into the vacuum column of a TEM microscope (JEOL 1400 operating at 120 kV) where it was maintained at liquid nitrogen temperature thanks to a cryo holder (Gatan). The magnification was selected between 3000× and 40000×, and images were recorded with a 2k-2k Ultrascan camera (Gatan).

## 3. Results and Discussion
### 3.1. Silicone dispersion in presence of TEQ

The development of a novel silicone-oil containing softener was first attempted by using TEQ double-chain surfactant for stabilizing silicone oil droplets before incorporation into the softener formulation (Scheme 1a). A 20 wt. % AMPS emulsion was prepared by mixing AMPS with 2 wt. % TEQ into an Ultraturax system (AMPS$_{TEQ}$). At this concentration TEQ is assembled into multivesicular or unilamellar vesicles of wide size distribution (0.1 µm – 3µm), as shown by cryo-TEM and optical microscopy given at Figure 1d and DLS data shown elsewhere. [1] However, the vesicles break to smaller ones by applying external stresses such as sonication, filtration, or Ultraturax homogenizer. [52] Therefore, we assume that into the emulsion the majority of TEQ vesicles are smaller than the vesicles presented in Figure 1d. The characterization of the AMPS$_{TEQ}$ dispersion (20 wt. %) is summarized in Figure 2. The size distribution (hydrodynamic diameter, $D_H$) obtained for the AMPS emulsion at the concentration at which is formulated c= 20 wt. % and at a diluted one, c= 2 wt. %, are given in Figure 2a. Two main size populations, at 0.1 and 1.5 µm, are present for both concentrations. Here, the size concerns mainly AMPS droplets and few possible TEQ free vesicles as TEQ concentration is ten times lower than AMPS and the majority of TEQ should be somehow linked to the AMPS droplets as it will be discussed below. Besides, oil droplets exhibit higher scattering intensity compared to vesicles. Electrophoretic mobility measurements (Figure 2b) show positive charge for AMPS emulsion ($\zeta$ = + 53 mV). The zeta potential obtained for TEQ 0.01 wt. % is reported at $\zeta$ = + 59 mV indicating no zeta potential modifications provoked by mixing AMPS with TEQ, while their positive charges favors the adsorption onto the negative cellulosic fabrics.

Phase contrast optical microscopy experiments were performed to visualize the emulsion structure. Figure 2c displays a 40× magnification image of the dispersion at $c$ = 20 wt. %. The formulation is homogeneous while highly contrasted and densely packed spherical objects of average size 800 nm which undergo Brownian motion (**Supporting Information, movie No #1**) are observed. These objects are identified as silicone oil droplets due to their significant number and spherical shape. Moreover, different contrast and size distribution is found compared to the TEQ vesicles. [1] Note here that the observation of white and black objects is assigned to differences in focus. To confirm the existence of silicone droplets we used bright field optical microscopy studies. As vesicles encapsulate water, they exhibit very low contrast and are not well captured when bright field mode is applied, in contrast to oil. Hence, the densely packed spherical objects revealed in Figure 2d are attributed to AMPS oil droplets. Interestingly, on the surface of the larger ones, some white or black spheres (depending on the focus) are often observed. Those adhered spheres are indicated by yellow arrows and are better visualized at higher magnification in the inset of this figure. To get a better insight, we diluted the dispersion at $c$= 0.2 wt. %. As revealed in Figure 2e, various types of structures are now visible. Droplets which seem like they have 'holes' on their surface (red arrows). This particular form could be explained by the partial engulfment of vesicles onto the oil surface. As vesicles are not visible in the bright field mode, they appear as spherical 'holes'. The partial vesicles engulfment onto oil surfaces has been already reported [43,44] (morphology shown in Figure 2f) while a mechanism suggesting that the bilayer is ruptured to a monolayer adhered to the oil while the rest keeps its bilayer form has been proposed. [45]

Additionally, different structures of spherical objects onto the surface (Figures. 2g-i) are found. Here the objects can be adhered on the surface or they are completely engulfed into the droplet (appearing to be dark and not as holes). A better visualization is offered by the **movie #2** in **Supporting Information** where we can distinguish smaller spheres which are moving although linked to the surface of the larger spherical objects as well as 'holes' derived by the partial engulfment of vesicles onto the oil droplet. Similar structures have been already mentioned for vesicle/oil interaction. [43] The stabilization of oil droplets in water by vesicles has been scarcely mentioned in the literature while it is explained either by the formation of Pickering emulsions [46] (form I, and III in Figure 2j) or by the adsorption of a monolayer onto the droplet (form IV). [44] In our case we have probably obtained this latter structure too but it cannot be visualized at this scale. Our formulation is a mixture of various possible structures (Figure 2j) which permit the stabilization of the AMPS into the water in a small size. On the contrary, as expected, in absence of TEQ surfactant a 20 wt. % AMPS dispersion in water prepared by Ultra-Turrax resulted in a two-phase dispersion.

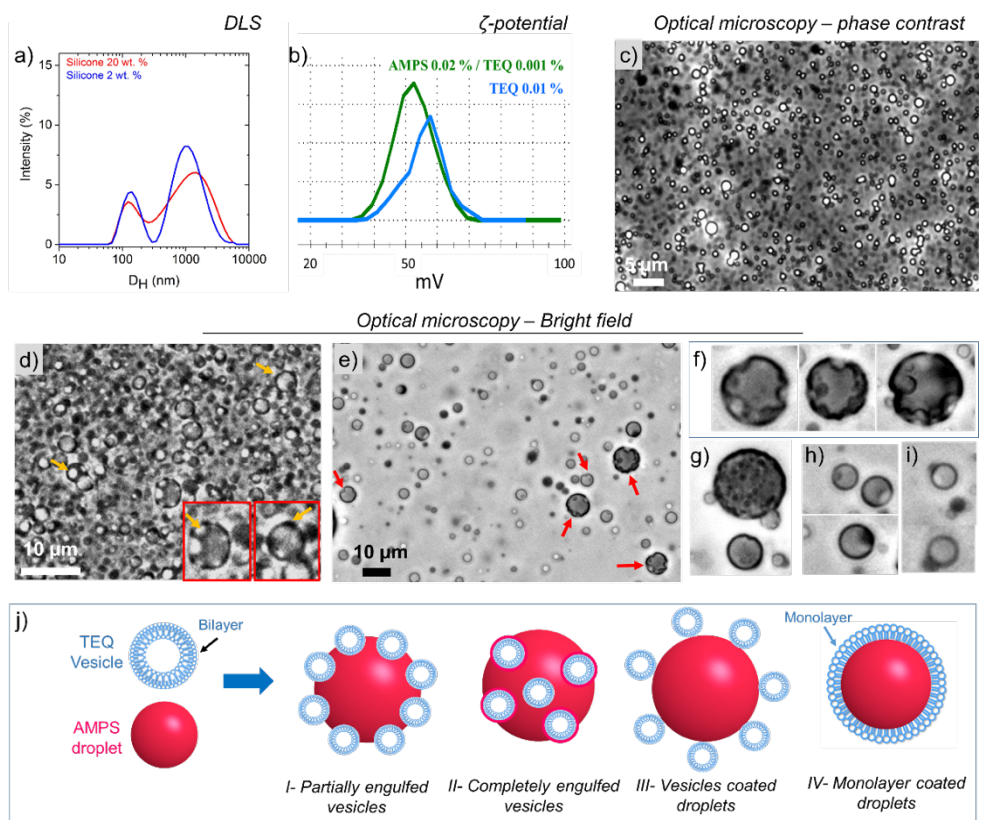

***Figure 2**. (a) Size distribution obtained by DLS for 20 wt. % and 2 wt. % AMPS$_{TEQ}$ dispersions, (b) zeta potential results for the formulations TEQ 0.01 wt. %. and AMPS$_{TEQ}$ diluted at 0.02 wt. % (c) representative phase contrast optical microscopy image × 40 of 20 wt. % AMPS$_{TEQ}$, (d) same in bright field mode, (e) bright field optical microscopy image × 40 of the AMPS$_{TEQ}$ diluted at 0.2 wt. %, (d) characteristic structures obtained for the 0.2 wt. % AMPS$_{TEQ}$ (bright field) (j) schematic representation of the possible structures obtained for AMPS$_{TEQ}$.*

### 3.2. Incorporation of AMPS$_{TEQ}$ into the surfactant formulation

Before incorporating the silicon emulsion into the guars-surfactant softener, we checked the effect of the AMPS$_{TEQ}$ on the TEQ vesicles upon its incorporation in a 6 wt. % TEQ formulation. The concentration of AMPS$_{TEQ}$ into the 6 wt. % TEQ was adjusted at 1 wt. %. Optical microscopy characterization (phase contrast) revealed numerous spherical droplets of diameter D < 1 µm (yellow arrows) distributed among numerous TEQ vesicles (green arrows) (Figure 3a). The larger vesicles exhibit facets, as discussed in section 2. In the inset of Figure 3a, these well contrasted spherical droplets attributed to AMPS, are better visualized (yellow arrows). Their contrast is different from the spherical vesicles and are not detected in Figure 3b, where a representative image of TEQ 6 wt. % without AMPS is exhibited. By comparing Figure 3a with 3b no difference on the vesicles size and morphology is observed, indicating that the structure of the TEQ formulation is not influenced by AMPS$_{TEQ}$ insertion, at least at this observation scale. TEQ dispersions at 6 wt. % and diluted at 0.1 wt. %, in presence (solid line) and in absence of AMPS (dash-dot) were analyzed by DLS and are presented in Figure 3c. Two size populations are obtained for the dispersions studied. No significant impact of AMPS on the TEQ size is recorded, whatever the concentration. A better insight on the AMPS droplets into the TEQ formulation was attempted by bright field microscopy (Figure 3d) where the vesicles cannot be well observed. Therefore, we conclude that the spherical well contrasted objects are AMPS droplets and that they are homogeneously distributed into the formulation. The size distribution obtained from this image is peaked at 700 nm (Figure 3e). Dilution at *c*= 0.2 wt. % (Figure 3f) permits us to distinguish partially or entirely engulfed vesicles to the silicone droplets as those analyzed in Figure 2. These findings show that the AMPS droplets structure do not change upon incorporation to the TEQ formulation.

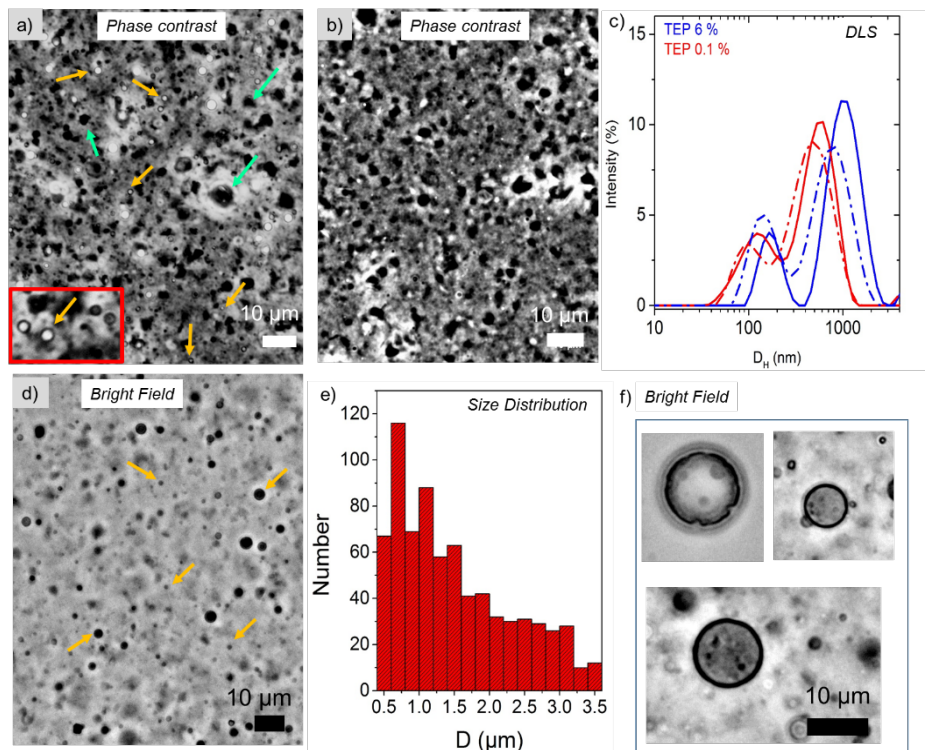

**Figure 3**. *(a) Representative optical microscopy image of TEP 6 wt. % in presence of AMPS$_{TEQ}$ 1 wt. % at ×40, inset: AMPS droplets (yellow arrows) at higher magnification, (b) representative optical microscopy image of TEQ 6 wt. %, (c) Size distribution of TEQ 6 wt. % and 0.1 wt. % in presence (solid line) and in absence (dash dot line) of AMPS$_{TEQ}$, obtained by DLS, (d) bright field optical microscopy image of TEP 6 wt. % in presence of AMPS$_{TEQ}$ 1 wt. % at ×40 and (e) associated size distribution, (f) higher magnification bright field optical microscopy images of partially and entirely engulfed vesicles by AMPS droplets, $c_{TEQ}$= 0.2 wt. %.*

The AMPS$_{TEQ}$ emulsion was then incorporated into the guars/TEQ containing softener resulting in the novel formulation abbreviated as SOFT 1 (Scheme 1a). The structure of SOFT 1 consisted by TEQ 6 wt. %/AMPS$_{TEQ}$ 1 wt. %/C-Guar 0.2 wt. %/HP-Guar 0.4 wt. % was examined by phase contrast optical microscopy. Spherical well contrasted objects related to the AMPS droplets are well distributed in the formulation as indicated by the red arrows in Figure 4a. The appearance of the formulation is identical to the corresponding formulation [4] in absence of AMPS$_{TEQ}$ as shown by optical microscopy image provided in the **Supporting Information, Figure S2**. There, similar to Figure 4a, a micron sized texture (assigned to the guars/vesicles interaction) [4] instead of faceted and spherical vesicles is identified. The silicone oil droplets are better distinguished (red arrows) by bright field optical microscopy (Figure 4b) since TEQ and guars cannot be recorded at this mode. The AMPS droplets are presented at a higher magnification in the inset of Figure 4b and their size distribution is peaked at 1.5 µm (Figure 4c).

Cryo-TEM studies of oil/vesicles systems for pharmaceutical or food applications have been rarely reported [44,53] while silicone oil/vesicles cryo-TEM characterization has not been mentioned so far. We proceeded in cryo-TEM experiments to identify the influence of AMPS on the vesicles at the nano-metric scale. The studies revealed the co-existence of very small dark spheres ascribed to AMPS droplets (due to their higher electron scattering) indicated by the blue arrows in Figures. 4 d-g along with multi-vesicular vesicles. The size of these droplets is peaked at 25 nm as shown in Figure 4i. The droplets are found between the vesicles. The vesicles structure and the bilayer thickness remain stable after the silicone oil insertion in the formulation. This is confirmed by comparing Figures. 4d-g with a cryo-TEM image of the formulation before AMPS addition provided in the **Supporting Information, Figure 2b**. In contrast to optical microscopy studies, with cryo-TEM we identify only small AMPS droplets assuming that are coated by a monolayer (form IV, Figure 2i) as the droplets cannot be stabilized in such a small size without an emulsifier. Those findings ascertain that AMPS can be incorporated into the guars/esterquat containing softener without causing changes in the structure of the formulation. The selected strategy for its insertion resulted in an appropriate silicone distribution between the vesicles while the size of the droplets varies between 0.01 and 2 μm detected by cryo-TEM and optical microscopy.

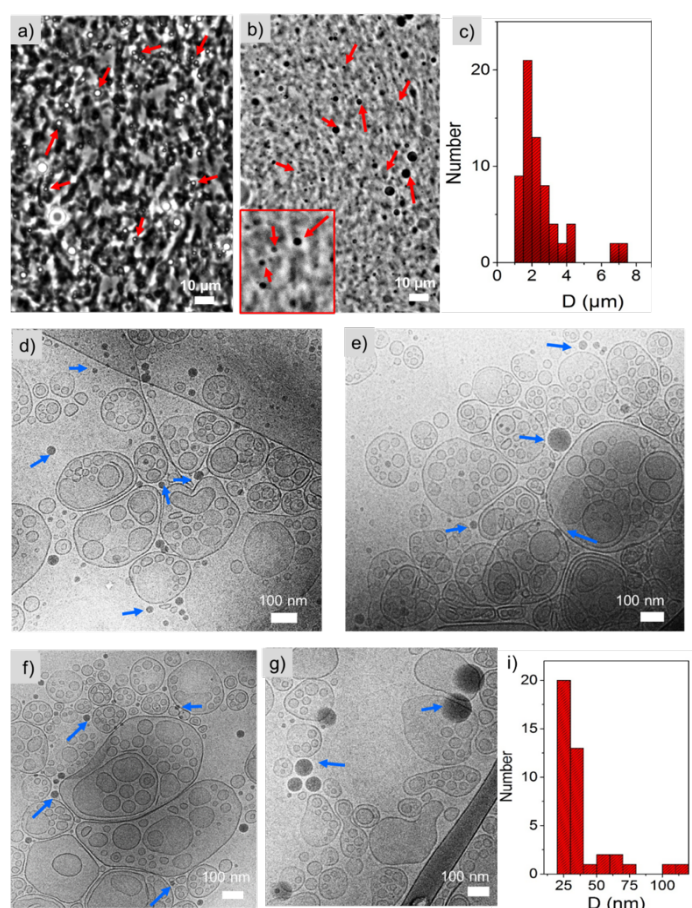

*Figure 4*. Representative optical microscopy images of SOFT 1 (TEP 6%/AMPS$_{TEQ}$ 1%/C-Guar 0.2%/HP-Guar 0.4 %) in (a) phase contrast, (b) bright field mode at ×20. The AMPS droplets are indicated by the red arrows. (c) AMPS droplets size distribution associated to (b). (d-g) cryo-TEM representative images of the SOFT 1 diluted at TEQ 1 wt. %. AMPS droplets are indicated by the blue arrows. (i) AMPS droplets size distribution obtained for the figures (d-g).

### 3.3. Silicone emulsification in presence of HP-Guar – incorporation into the surfactant formulation

An alternative strategy based on the use of HP-guar for stabilizing AMPS oil droplets before incorporation into the guars/esterquat softener is proposed here and is schematically represented in Scheme 1b. In contrast to the approach followed for SOFT 1 where a concentrated stock solution of AMPS stabilized by TEQ vesicles was used, here the AMPS and HP-Guar concentration correspond to their concentration in the final formulation (SOFT 2). Thus the emulsion is used as it is in the following steps. Polysaccharides have been rarely reported as oil emulsifiers while their emulsifying ability is usually strengthened by proteins. However, despite the existing debate on their emulsifying ability, it has been proven that even in absence of proteins, hydrocolloids like gums can be effective for stabilizing oil droplets. [48,54] Here, both C-Guar and HP-Guar were examined as for their emulsifying properties. C-Guar was found to be inappropriate for such a process. On the contrary, AMPS oil in presence of 0.4 wt. % HP-Guar allowed for the formation of small uniform oil droplets. The obtained emulsion (AMPS$_{HP-Guar}$) was characterized by optical microscopy. A homogeneous dispersion of droplets exhibiting diameter D < 1 μm was obtained and it is exhibited in Figure 5a.

To ensure that those droplets correspond to AMPS and not to HP-Guars colloidal particles, fluorescent microscopy was also applied. Prior to that, a fluorescent dye known to interact with lipid surfaces, PKH67 was mixed with the emulsion assuming that the hydrophobic nature of oil would permit the dye insertion. [55] Indeed, as seen in Figure 5b numerous bright spots are recorded. Their size is in good agreement with that obtained from optical microscopy. In parallel, a negative experiment was conducted on HP-Guar aqueous dispersion after mixing with the dye and no fluorescent signal was recorded. A movie (**movie #3 Supporting Information**) ascertains their movement as well as the signal loss upon time due to bleaching. Those results demonstrate that HP-Guar can stabilize/emulsify silicone oil droplets at these experimental conditions. The increased viscosity that the guar imparts to the aqueous continuous phase probably helps to limit the mobility of the dispersed oil phase via the formation of a structural-mechanical barrier that results in an improved kinetic stability of the emulsion. This statement is further supported by a movie given in the **movie #4 Supporting Information** showing that the droplets are moving slowly compared to the mobility they exhibit at the same concentration stabilized by TEQ (**movie #1 and movie #2 Supporting Information**.) The dynamic rheological properties of HP-Guar have been measured and are provided in the **Supporting Information, Figure S3**. Besides, it has been claimed in the literature that polysaccharides containing a hydrophobic group on the chain can be adhered on the oil surface. [49] It could be though possible that apart from the viscosity gel-like mechanical properties, the hydroxypropyl group onto the polymer chain contributes to the droplets stabilization. The polysaccharide concentration impact on the emulsion properties was examined by doubling its concentration. In Figure 5c, a representative optical microscopy image of an emulsion prepared with 0.8 wt. % HP-Guar and 1 wt. % AMPS is presented. Although phase separation was prevented, the formulated droplets are characterized by a large size distribution varying between 1 and 20 μm. The very high viscosity of the guar at this concentration did not allow for a homogeneous mixture hindering applied stress from Ultra-Turrax to influence the droplets. Finally, it should be pointed that although the method proposed here results into AMPS emulsion with good size distribution, it concerns low AMPS concentration. 20 wt. % AMPS in presence of 0.4 wt. % HP-Guars resulted in inhomogeneous droplets (**Supporting Information, Figure S4**) indicating that the polymer was not sufficient to stabilize the formed droplets. However, this does not impose problems in our case as we desire AMPS 1 wt. %. The emulsion is used directly as it is without dilution for the softener formulations.

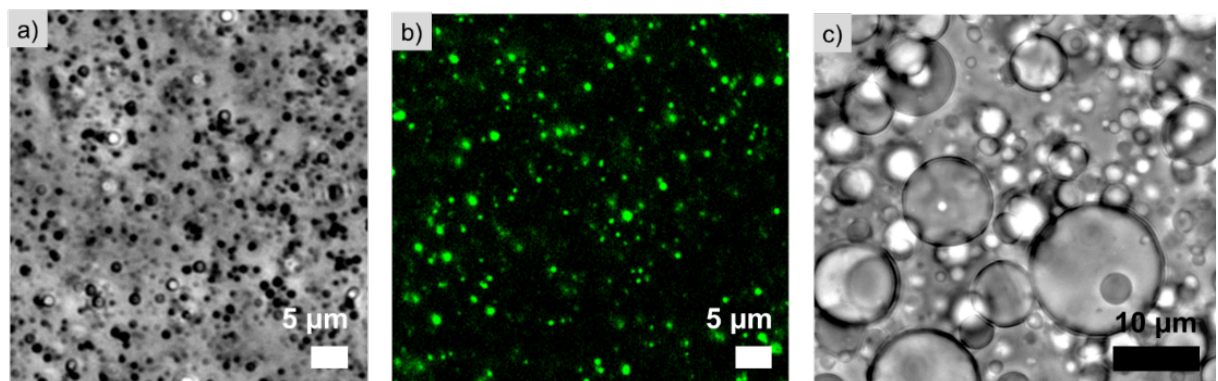

**Figure 5.** *Representative (a) optical and (b) fluorescence microscopy images of AMPS % 1 wt. stabilized by HP-Guar 0.2 % wt. (c) Representative optical microscopy image of AMPS 1 wt. % stabilized by HP-Guars 0.8 wt. %.*

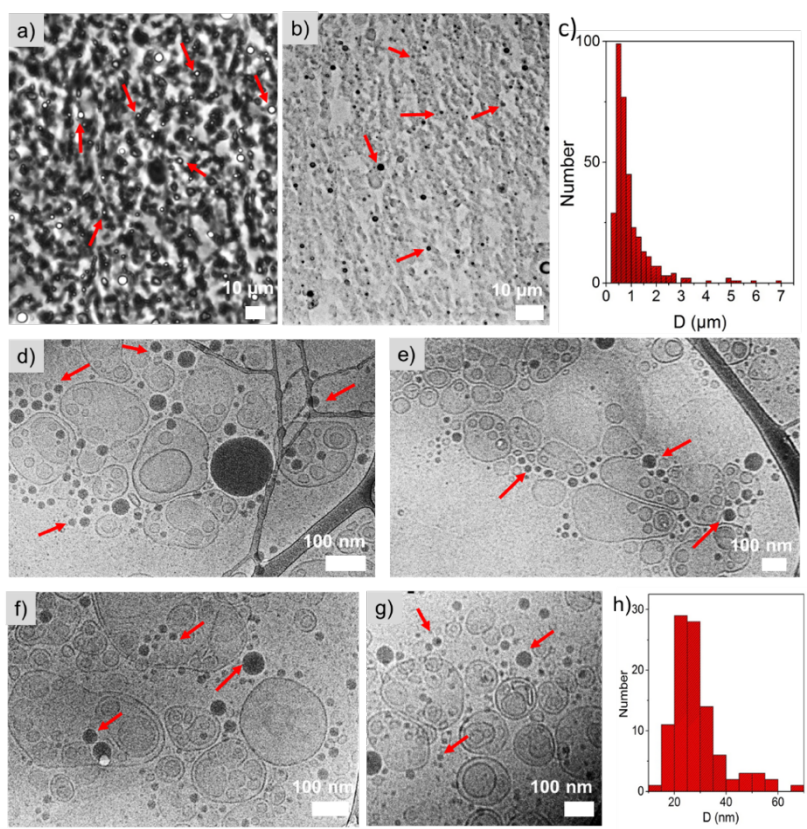

**Figure 6.** *Representative optical microscopy images of SOFT 2: TEP 6%/AMPS$_{HP\text{-}Guar}$ 1%/C-Guar 0.2%/HP-Guar 0.4 % at ×20 in (a) phase contrast, (b) bright field mode. The AMPS droplets are indicated by the red arrows. (c) AMPS droplets size distribution associated to (b). (d-g) cryo-TEM representative images of the SOFT 2 diluted at TEQ 1 wt. %. AMPS droplets are indicated by the red arrows. (h) AMPS droplets size distribution obtained for the figures (d-g).*

The resulting softener (SOFT 2) after addition of C-Guar and TEQ (Scheme 1b) was characterized by optical and cryo-TEM studies. The results are summarized in Figure 6. Optical microscopy in both phase contrast and bright field mode (Figures 6 a and b respectively) reveal an identical structure to SOFT1 (Figure 4) and a bit smaller size (Figure 4c). The size distribution is peaked at 500 nm. Cryo-TEM analysis showed co-existence of multi-vesicular vesicles along with small (< 80 nm) dark spheres ascribed to AMPS droplets. Those findings witness the appropriate incorporation of $AMPS_{HP-Guar}$ oil into the softener enabled by HP-Guar emulsifying effect.

### 3.4. Long-term stability of the formulations

The stability of the prepared emulsions and softeners formulations was followed by visual observation and optical microscopy. In Figure 7, optical microscopy images for AMPS emulsified by TEQ (a) or HP-Guar (b) as well as SOFT 1 (c) and SOFT 2 (d) are exhibited at a certain time after their preparation. $AMPS_{TEQ}$ is stable 15 days after its formulation with naked eye observation but by microscopy, larger droplets (red arrows) and some aggregates are detected. After 25 days, the emulsion starts to phase separate and cannot be used. On the contrary, $AMPS_{HP-Guar}$ is very stable even after 2 months. In Figure 7b, the microscopy image is taken 2 months after its preparation. Although some larger droplets (red arrows) are monitored, the emulsion and the droplets distribution and size remain unmodified. Pertaining into the guars containing softeners, both of them, SOFT 1 and 2 are found to be stable five months after their formulation. Both their texture and the droplets size (red arrows) do not undergo any change. Regarding the different stability observed for the $AMPS_{TEQ}$ and SOFT 1, it is suggested that the numerous vesicles along with the guars gel-like structure in the formulation contribute to the longer stability of the oil droplets.

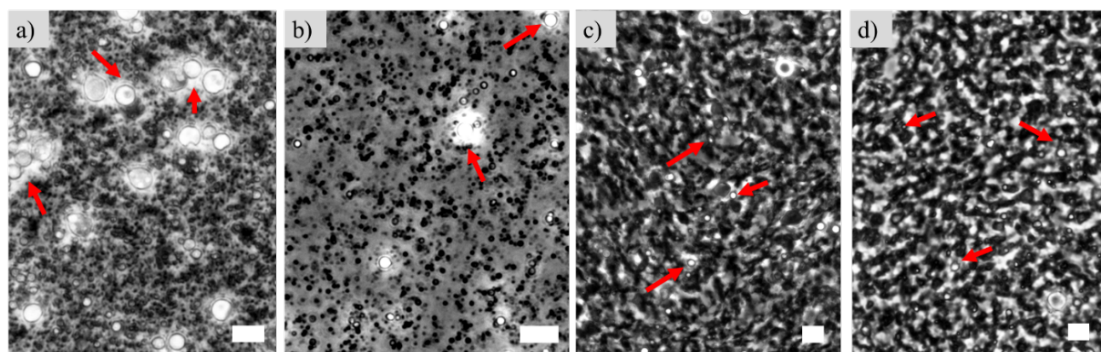

**Figure 7**: Representative phase contrast optical microscopy images of (a) $AMPS_{TEQ}$, (b) $AMPS_{HP-GUARr}$, (c) SOFT 1:TEP 6%/$AMPS_{TEQ}$1%/C-Guar 0.2%/HP-Guar 0.4 % and (d) SOFT 2: TEP 6%/$AMPS_{HP-GUAR}$ /C-Guar 0.2%/HP-Guar 0.4 %. The scale bars are 10 μm.

### 4. Conclusions

Esterquat surfactant and amino-modified silicone oil, the active ingredients of two fabric softener classes, household and textile finishing, were successfully combined into a novel formulation. Minor quantities of guar polymers (C-guar and HP-guar) were also added as they were shown previously to enable the reduced esterquat content (6 wt. %) compared to the commercial concentrated conditioners (> 10 wt. %). The silicone oil incorporation into the industrial softener formulation was achieved following two different novel strategies: instead of adding a surfactant for silicone emulsification, the softener components, either esterquat vesicles or HP-Guars, were used to stabilize the silicone droplets. In case of using the surfactant vesicles, a concentrated stock silicon emulsion was prepared and was then introduced into the softener preparation in adequate quantities. Various possible structures, vesicles partial engulfment, adsorption, inclusion or surfactant monolayer adsorption on the oil droplets permit the oil dispersion. When using HP-Guar, the emulsion was prepared at the final desired guar and silicone concentration and was further used as it is for the softener preparation. In both cases, sub-micron and nano-droplets were identified by optical microscopy and cryo-TEM respectively. The co-existence of both size distributions is considered to enable simultaneous external and internal adsorption on the cotton fibers. The silicon oil droplets were shown to be well distributed between the vesicles while the silicone oil addition did not influence the vesicles structure, the bilayer thickness and the physical-chemical properties of the formulation. Both strategies resulted in a formulation which exhibits good long-term stability. The advantage of TEQ as dispersant medium is that the silicone concentration at the final formulation can be modified because a stock solution is used. On the other hand, the use HP-Guar as dispersant medium is restricted to specific silicone concentrations but its preparation is included in the softener preparation avoiding additional steps. The proposed methodologies can be further for the development of new softener formulations of various compositions.

## ACKNOWLEDGEMENTS

Annie Vacher (Solvay) and Marc Aurieu (Solvay) are acknowledged for the cryo-TEM experiments. We thank Imane Boucenna for providing us the Ultra-Turrax homogenizer. This research was funded by Solvay.

## ABBREVIATIONS

| | |
|---|---|
| DLS | Dynamic light scattering |
| Cryo-TEM | Cryogenic transmission electron microscopy |
| AMPS | Aminopropylmethylsiloxane – dimethylsiloxane copolymer |
| C-Guar | Cationic guar gum |
| HP-Guar | Hydroxypropyl guar gum |
| TEQ | Esterquat surfactant ethanaminium, 2-hydroxyN,N-bis(2-hydroxyethyl)-N-methyl-ester |
| $AMPS_{TEQ}$ | TEQ emulsified AMPS |
| $AMPS_{HP-Guar}$ | HP-Guar emulsified AMPS |
| SOFT 1 | Softener 1 |
| SOFT 2 | Softener 2 |

**Supporting Information**

Dynamic light scattering characterization of C-Guar and HP-Guar; Optical microscopy and cryo-TEM characterization of the reference softener; Guar linear rheology; $AMPS_{HP-Guar}$ 20 wt. % optical microscopy characterization; Movie #1: Phase contrast optical microscopy of $AMPS_{TEQ}$ 20 wt. %; Movie #2: Bright field optical microscopy of $AMPS_{TEQ}$ diluted at 0.2 wt. %; Movie #3: Fluorescent microscopy characterization of 1 wt. % $AMPS_{HP-Guar}$; Movie #4: Phase-contrast optical microscopy of 1 wt. % $AMPS_{HP-Guar.}$

# Graphical abstract

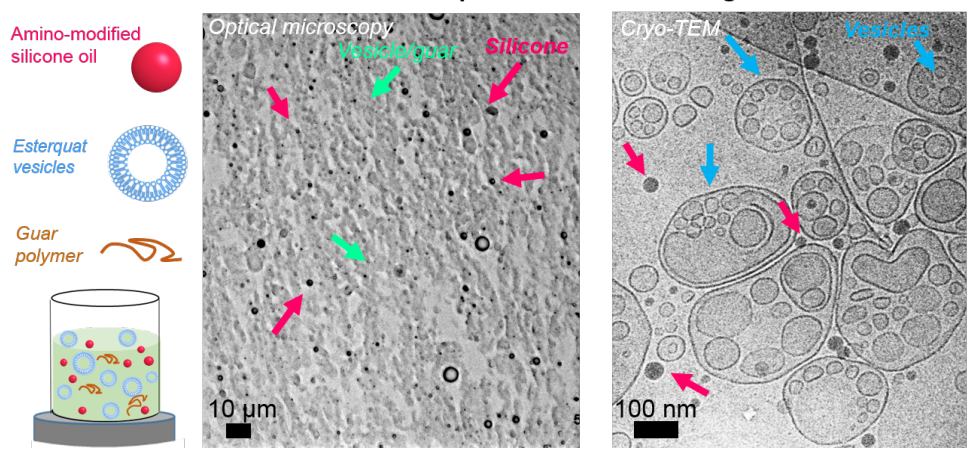

*Amino-modified silicone oil incorporation in a vesicles/guars formulation*